# How do financial variables impact public debt growth in China? An empirical study based on Markov regime-switching model

*Tianbao Zhou[1], Zhixin Liu[1], Yingying Xu[2*]*

1 School of Economics and Management, Beihang University, Beijing, China

2 School of Humanities and Social Science (School of Public Administration), Beihang University, Beijing, China

*Corresponding author: E-mail: yingxu21@buaa.edu.cn

**Abstract:** The deep financial turmoil in China caused by the COVID-19 pandemic has exacerbated fiscal shocks and soaring public debt levels, which raises concerns about the stability and sustainability of China's public debt growth in the future. This paper employs the Markov regime-switching model with time-varying transition probability (TVTP-MS) to investigate the growth pattern of China's public debt and the impact of financial variables such as credit, house prices and stock prices on the growth of public debt. We identify two distinct regimes of China's public debt, i.e., the surge regime with high growth rate and high volatility and the steady regime with low growth rate and low volatility. The main results are twofold. On the one hand, an increase in the growth rate of the financial variables helps to moderate the growth rate of public debt, whereas the effects differ between the two regimes. More specifically, the impacts of credit and house prices are significant in the surge regime, whereas stock prices affect public debt growth significantly in the steady regime. On the other hand, a higher growth rate of financial variables also increases the probability of public debt either staying in or switching to the steady regime. These findings highlight the necessity of aligning financial adjustments with the prevailing public debt regime when developing sustainable fiscal policies.

**Keywords:** Public debt; Credit; House prices; Stock prices; TVTP-MS;

**JEL Classification:** E32, E44, E60, H60, H63

# 1 Introduction

The outbreak of the COVID-19 pandemic in 2020 had a heavy strike on the world economy, leading to another global recession and financial turmoil ever since the financial crisis in 2008 (Goodell, 2020; The World Bank, 2020; Zhang et al., 2020). It also impacted the fiscal balance and public debt stability in China, triggering a new wave of public debt increase. As early as the surge in public debt worldwide after the 2008 financial crisis, many studies began to pay attention to the role of financial markets in shaping macroeconomic operations, government finance and public debt growth (e.g., Borio, 2014; Jordá et al., 2016; Reinhart and Rogoff, 2009). For instance, with the deepening financial liberalization, today's financial markets are closely linked to the real economy and affect fiscal revenue and public debt dynamics through various factors such as government taxation, personal consumption, and household wealth (Tagkalakis, 2011, 2012), thereby broadening the factors for traditional debt management and increasing its complexity. Given the rapid growth of China's public debt and financial turmoil in the post-COVID era, policymakers should not only consider macroeconomic factors, but also closely monitor changes in financial markets. In this context, identifying the growth patterns of China's public debt and discovering the heterogeneous impacts of financial variables on the growth of China's public debt help to formulate targeted financial policies under different debt conditions, which will collaboratively cope with the surge of public debt caused by external shocks and improve the future sustainability of China's public debt.

Indeed, the Chinese government has implemented a series of expansionary fiscal policies to stimulate economic recovery in the post-COVID era, such as large increases in budget expenditures, issuance of special government bonds, and reductions in corporate taxes and fees (e.g., Xu and Wei, 2021; Zhao, 2020). According to the Bank for International Settlements (BIS), the gross nominal public sector debt-to-GDP ratio in China had surpassed 70% by the end of 2020 and continued to rise in the following years. In addition, the downturn in financial markets has significantly exacerbated fiscal deficits and the rapid growth of China's public debt. Disruptions in production, lockdowns in the city, and trading restrictions also resulted in prolonged sluggish investments and credit, and falling asset prices, which has worsened corporate balance sheets, led to more credit defaults for commercial banks, and further constrained future investment activities (Rizwan et al., 2020). Meanwhile, the inactive market trading also reduced transaction-related tax of the government and net-worth household asset values, thereby suppressing consumption through the "reverse wealth effect" and lowering consumption tax revenues. These facts have compelled the government to maintain expansionary fiscal policies and issue more debt over longer periods, sparking concerns about the sustainability and risks of China's public debt (Cueastas and Regis, 2018; Hao et al., 2022). As proposed by the financial accelerator theory, financial variables should not be treated independently, as they may amplify the impact of external shocks on the cyclical fluctuations of the real economy (Bernanke et al., 1999), and affect the effectiveness of fiscal policies (Borsi, 2018; Soederhuizen et al., 2023). By adjusting certain financial variables, the government can better-moderate the rapid growth of China's public debt, as the stable growth of public debt helps to reduce the risk of future debt defaults and facilitate economic recovery after the pandemic.

The evolution and development of public debt and financial markets in China have distinct characteristics. First, China's rapid economic growth over the past three decades is attributed much to its unique land finance system. Local government debt is an important financing channel for Chinese local governments (Lin, 2010), and the windfall revenue brought by asset price booms further strengthens the government's exploitation and development of land assets, eventually forming a "land-infrastructure-leverage" strategy in China's urban development (Tsui, 2011). As land is often used as collateral for bank loans, the downturn in China's asset price markets and the decline in land premiums will then pose a negative impact on land finance, which may lead to more debt insurance as a supplement to the government's income, resulting in public debt expansions and greater debt risks (Cheng et al, 2022). Therefore, changes in asset prices are the main factor affecting the growth of China's public debt. Second, unlike most developed economies, there are many state-owned banks and policy banks in China's financial system, and the credit market serves as a key channel for public debt issuance. In particular, since China implemented a series of economic stimulus measures in 2009 in response to the global financial crisis, off-balance sheet local government financing vehicles have become the main units to borrow from banks, with many municipal corporate bonds implicitly guaranteed by local governments (Chen et al., 2020b). This feature strengthens the link between China's credit markets and public debt. Third, real estate loans account for a large proportion of China's commercial bank loans (Zhang et al., 2014). Falling house prices will significantly increase mortgage defaults, and credit risks and even threaten the financial system stability (Chen and Wen, 2017; Su et al., 2021). Since China's real estate policy is subject to direct government intervention, there is also a stronger cross-market linkage between the asset price market, credit market, and government finance. In addition, China's stock market is also very special in the world, such as the "T+1" trading and settlement system, the minimum requirements for investors in different sectors, and the regulation of stamp duty. This may affect investors' preference for investment types and market liquidity, being another factor affecting fiscal policies and government finance.

Given that many recent studies focus on the macroeconomic factors and fiscal consolidation of public debt in advanced economies (e.g., Cherif and Hasanov, 2018; Fukunaga et al., 2021; Georgantas et al., 2023), there is a lack of identification of the growth pattern of China's public debt and its financial factors. Investigating China's sample will deepen the understanding of public debt development and contribute to the existing literature empirically. The main contributions of this paper are as follows. First, to the best of our knowledge, this paper is the first to identify the regimes of China's public debt by the Markov regime-switching model with time-varying transition probability (TVTP-MS) based on a nonlinear framework. Our sample covers the post-COVID period in China, providing insights into how those crises impact public debt growth and policy responses over time. Second, this paper investigates the impacts of several financial variables on the growth rate of public debt under different regimes, and how they affect the transition probability of public debt between regimes. According to the empirical results, we identify that China's public debt shows high growth rate and high volatility in the periods of external shocks and crises, i.e., the surge regime, and low growth rate and low volatility in the rest of the periods，i.e., the steady regime. We find that a higher growth rate of credit and house prices reduces the growth rate of public debt in the surge regime, whereas an increase in the growth rate of stock prices reduces the growth rate of public debt in the steady regime. We also confirm that a higher growth rate of the financial variables also increases the probability of public debt either staying in or switching to the steady regime. In summary, the increase in the growth rate of financial variables helps to moderate the growth rate of China's public debt and facilitate public debt to grow in a more sustainable way. Our results reveal the characteristics of the growth pattern of China's public debt, supporting the literature that advocates combining financial markets with the real economy when conducting macroeconomic research after the global financial crisis (e.g., Bernanke et al., 1999; Kiyotaki and Moore, 1997; Tagkalakis, 2011, 2012). It highlights the importance of the linkage between the financial sector and the fiscal sector in managing China's public debt and provides insights for the development of targeted financial policies to improve the sustainability of public debt and reduce debt pressure.

The remainder of this paper is as follows: Section 2 is the literature review. Section 3 introduces the model, the data, and the stylized facts of China's public debt. Section 4 is the empirical results. Section 5 concludes.

## 2 Literature Review

*2.1 The nexus between public debt and private debt*

The first strand of literature investigates the nexus between public debt and private debt and reveals that changes in private debt can meanwhile influence the effectiveness of fiscal policy and the growth of public debt. Jorda et al. (2016) conclude that high public debt levels exacerbate the cost of private sector deleveraging after an adverse financial shock, resulting in a deeper recession. Some studies on financial crises suggest that high leverage and excessive credit boom aggravate post-crisis recession, fiscal deterioration and debt space compression (e.g., Borio, 2014; Jorda et al., 2016; Reinhart and Rogoff, 2009). Batini et al. (2019) find that after a financial shock, recessions are milder and the stability of public debt is better controlled when governments lend directly to those households and corporate facing borrowing restrain. Andrés et al. (2020) apply a DSGE model and discuss the channels between private and public debt reduction. Moreover, private debt also affects public debt growth by changing the size of the fiscal multiplier. For example, based on the US sample, Bernardini and Peersman (2018) find that government spending multipliers are much larger during periods of private debt overhang, and more government purchases can also reduce the public debt-to-GDP ratio. Borsi (2018) and Eggertsson and Krugman (2012) also confirm similar conclusions through other models.

*2.2 The impact of house price and stock price on public debt*

The second strand of literature reveals the impact of house prices and stock prices on the dynamics of public debt. In the seminal paper of Kiyotaki and Moore (1997), they mention that durable assets play a dual role, i.e., as factors of production and as collateral for loans. Thus, if the value of collateral changes, the demand shock can be amplified, which in turn affects government fiscal behavior and public debt issuance. The link between asset prices and public debt has also been studied in the post-crisis period. Tagkalakis (2011, 2012) finds that rising asset prices have a positive impact on the primary balance of payments, reflecting the increase of government revenue and the decrease of budget deficit, among which stock and house prices are key determinants. The literature also concludes that asset prices affect fiscal and public debt mainly through taxation. On the one hand, house prices and stock prices affect direct taxes on households and corporate through capital gains taxes. When the price rises, the market is active, and the government can also raise revenue through turnover taxes on property transactions and stamp duty on stocks. On the other hand, higher house prices and stock prices also raise expectations and consumption through the "wealth effect" and increase indirect taxes (Chen et al., 2020a; Lu et al., 2020; Wang et al., 2021). Tagkalakis (2011) also points out that if the collapse of asset price markets causes financial instability and leads to a negative feedback loop of economic activity, the government may adopt expansionary fiscal measures to avoid a comprehensive recession in the economy, resulting in a worsening of its budget position and a surge in public debt. Therefore, the changes in house prices and stock prices are also a non-negligible factor affecting the growth of public debt.

*2.3 Public debt dynamics under multiple regimes*

The last strand of regimes is based on the nonlinear framework. Luigi and Huber (2018) apply the threshold VAR model to determine the "high debt" regime and the "low debt" regime in the US and find that the effect of monetary policy is insignificant under the high debt regime. Besides, the transmission channels of monetary policy also show structural differences under different debt regimes. Owusu et al. (2022) employ the panel smooth transition regression model to identify "high debt" and "low debt" regimes, and examine the difference in the impact of fiscal shocks on public debt under different regimes. Different from Owusu et al. (2022), who use the levels of public debt-to-GDP ratio as the regime variable, Luigi and Huber (2018) analyze the growth rate of public debt-to-GDP ratio. There are also studies on public debt under different fiscal regimes. For instance, Chamorro-Narvaez and Zapata-Quimbayo (2024) use the MS model and the Bai-Perron structural change model and identify sustainable and unsustainable fiscal regimes in Colombia from 1980 to 2021. Under the two regimes, the authors demonstrate the asymmetric response of the government's primary balance to the increase in public debt. For other studies on public debt under different policy regimes, see Afonso et al. (2011), Afonso and Jalles (2017), Cassou et al. (2017), and Magazzino and Mutascu (2022). The above literature shows that the factors of public debt under different regimes are quite different, which demonstrates that the growth pattern and operation mechanism of public debt may also change according to different external economic conditions.

## 3 Methodology and data

*3.1 Model specifications*

This paper employs the TVTP-MS to identify two regimes of China's public debt and examines how financial variables impact public debt growth under different regimes and the transition probability. Hamilton (1989) first proposes the Markov regime-switching model to identify the US business cycle. Based on a nonlinear framework, the model is aimed to study the evolution characteristics and influencing factors of economic and financial series under different regimes. Since the transition probability of the model is constant, the model is also specifically referred to as the Markov regime-switching model with fixed transition probability (FTP-MS). Subsequently, Filardo (1994) and Diebold et al. (1999) extend the FTP-MS to the TVTP-MS. Compared with the FTP-MS, the TVTP-MS not only estimates the transition probability between regimes but also allows the probability to change with exogenous variables. Therefore, the TVTP-MS has been more widely used to study the formation mechanism, influencing factors, and time-varying characteristics of economic and financial sequences (e.g., Duprey and Klaus, 2022; Francis et al., 2021). According to Filardo (1994) and Diebold et al. (1999), this paper employs the TVTP-MS to investigate the impact of financial variables on public debt growth. The regression part of the model is as follows.

$$pd_t = \mu^s + \beta_1 y^s_{t-l_1} + \beta_2 ca^s_{t-l_2} + \beta_3 r^s_{t-l_3} + \beta_4 oil^s_{t-l_4} + \varepsilon^s_t \tag{1}$$

where, $u$ is the constant term, $\varepsilon$ is the residual term following $N \sim (0, \sigma^{2^s}_t)$. $pd$ is the growth rate of public debt[1], $y$ is the growth rate of industrial added value, $ca$ is the current account balance-to-GDP ratio, $r$ is the interest rate, $oil$ is the growth rate of the international crude oil price. Since this paper focuses on the short-term dynamics of China's public debt, the lags of the above variables (control variables) are determined by the minimum order that results in a significant coefficient. $s = [1, 2]$ is an unobservable regime variable following the first-order Markov chain. In the transition probability part of the model, the transition probability matrix is as follows.

---
[1] More specifically, $pd$ is the growth rate of the public debt-to-GDP ratio, which we will introduce in Section 3.2. For brevity, we refer public debt-to-GDP ratio as public debt. In addition, the use of public debt refers to the public debt-to-GDP ratio throughout the paper if there is no specific declaration.

$$\begin{pmatrix} p_{1\to 1} & p_{2\to 1} \\ p_{1\to 2} & p_{2\to 2} \end{pmatrix} \qquad (2)$$

$$p_{i\to j} = (s_t = j \,|\, s_{t-1} = i) \qquad (3)$$

$$p_{1\to 1} = \frac{exp(\alpha_0^1)}{1+exp(\alpha_0^1)} \qquad (4)$$

$$p_{2\to 2} = \frac{exp(\alpha_0^2)}{1+exp(\alpha_0^2)} \qquad (5)$$

where $\alpha_0^1$ and $\alpha_0^2$ are the constant terms in E.q. (4) and E.q. (5). $p_{i\to j}$ is the probability of switching to regime $j$ in the next period while currently under regime $i$. $i, j = 1, 2$ and satisfy the following equations.

$$p_{1\to 1} + p_{1\to 2} = 1 \qquad (6)$$

$$p_{2\to 2} + p_{2\to 1} = 1 \qquad (7)$$

Note that we call E.q. (1) the benchmark model as it only accounts for the four control variables, and neither E.q. (4) nor E.q. (5) involves any exogenous variables. Therefore, this is the FTP-MS. Next, we include financial variables in the regression part and the transition probability part respectively. By adding financial variables into E.q. (1), the regression part of the model is then as follows.

$$pb_t = \mu^s + \beta_1 y_{t-l_1}^s + \beta_2 ca_{t-l_2}^s + \beta_3 r_{t-l_3}^s + \beta_4 oil_{t-l_4}^s + \beta_5 fin_{t-l_5}^s + \varepsilon_t^s \qquad (8)$$

where $fin$ is the examined financial variable which can either proxies the growth rate of credit ($credit$), the growth rate of house prices ($house$) or the growth rate of stock prices ($stock$). When adding financial variables into E.q. (4) and E.q. (5), we have the TVTP-MS and the transition probability part of the model is as follows.

$$p_{i\to j} = (s_t = j \,|\, s_{t-1} = i, fin_{t-l_0}) \qquad (9)$$

$$p_{1\to 1}(fin_{t-l_0}) = \frac{exp(\alpha_0^1+\alpha_1^1 fin_{t-l_0})}{1+exp(\alpha_0^1+\alpha_1^1 fin_{t-l_0})} \qquad (10)$$

$$p_{2\to 2}(fin_{t-l_0}) = \frac{exp(\alpha_0^2+\alpha_1^2 fin_{t-l_0})}{1+exp(\alpha_0^2+\alpha_1^2 fin_{t-l_0})} \qquad (11)$$

where $i, j = 1, 2$. When $\alpha_1^1$ ($\alpha_1^2$) is positive, the probability of $p_{1\to 1}$ ($p_{2\to 2}$) increases with the increase of $fin_{t-l_0}$, and vice versa. $l_5$ and $l_0$ are the lags, determined by the AIC criterion. Besides, the transition probability still satisfies the equations below.

$$p_{1\to1}(fin_{t-l_0}) + p_{1\to2}(fin_{t-l_0}) = 1 \tag{12}$$

$$p_{2\to2}(fin_{t-l_0}) + p_{2\to1}(fin_{t-l_0}) = 1 \tag{13}$$

Therefore, we are able to reveal the complete characteristics of public debt regimes by only knowing $p_{1\to1}$ and $p_{2\to2}$.

*3.2 Data*

The sample of this paper is based on the quarterly data of China from 1998Q1-2023Q2, covering several major economic events in recent decades, such as the global financial crisis in 2008 and the COVID-19 epidemic in 2020. We use the gross nominal public sector debt-to-GDP ratio (public debt ratio) as the proxy for public debt (e.g., Antonini et al., 2013; Luigi and Huber, 2018). Moreover, credit, house prices and stock prices are widely used in the literature as proxies for financial cycles and financial markets (Claessens et al., 2012; Poghosyan, 2018; Shen et al., 2019), as they reflect the relationship between investment and savings, and the dynamics of the price market. Therefore, we use the gross nominal private sector debt-to-GDP ratio (credit ratio), commercial house prices, and the Shanghai Composite Index as proxies for credit, house prices and stock prices.

In addition, we also select four macroeconomic variables as the control variables of the model, i.e., industrial added value, current account balance-to-GDP ratio, interest rate, and the international crude oil price. Among them, the interest rate is proxied by the 30-day weighted average lending rate, and the international crude oil price is proxied by the international crude oil settlement price of the New York Stock Exchange. Given that economic growth and fiscal conditions affect government revenues and expenditures, interest rate affects government financing costs, and international factors also have a certain impact on the international balance of payments and domestic markets (Zhang, 2018), the inclusion of these control variables contributes to the reliability of the results. In this paper, all variables are in the growth rate series (log difference series) except for the current account balance-to-GDP ratio and the interest rate. The current account balance-to-GDP ratio and interest rate are in levels. The X-12 seasonal adjustment is conducted when the series shows significant seasonal fluctuations, all variables are in percentages. Public debt and credit data are from the BIS statistics (https://www.bis.org/statistics), whereas the rest of the variables are from the National Bureau of Statistics of China (https://data.stats.gov.cn/) and CEIC database (www.ceicdata.com).

*3.3 Descriptive analysis and stylized facts*

Table 1 presents the basic statistics of the variables. The results show that the average growth rate of public debt and three financial variables are close, around 1% to 2% per quarter. The growth rate of stock prices and house prices show large standard deviations and extreme values, which is consistent with the immature Chinese stock market (Liu et al., 2019) and the asset price fluctuations brought by China's real estate boom in the past two decades. In addition, the unit root test indicates that the growth rate series of public debt and financial variables are stationary.

**(Insert Table 1 here)**

Fig. 1 shows the growth rate of public debt, credit, house prices, and stock prices. There are four significant surges in public debt over the sample, i.e., in 2007, 2009, 2016 and 2020. These four debt surges are caused by either external shocks, major economic events or crises, having great impacts on the stability of China's public debt.

The first debt surge happened in 2007. In the second half of 2007, the subprime crisis began to engulf the US, leaving Chinese investments there exposed, including China's state foreign exchange reserves, domestic financial institutions, and some qualified domestic institutional investor (QDII) products. The crisis also reduced Chinese exports and stoked inflation, hindering economic growth. In response to the shock, China then launched an expansionary fiscal policy. In 2007, the Ministry of Finance issued 35 bonds worth 2.35 trillion yuan, an increase of 1.46 trillion yuan over the previous year. The public debt ratio increased significantly during this period, averaging about 3.3% per quarter The second debt surge occurred in 2009. With the impact of the global financial crisis continuing to expand, Greece's sovereign debt default led to the European debt crisis. The effect has spread from foreign exchange markets to capital and currency markets in many countries. Given China's stock market crash and the high domestic and external economic uncertainty, coupled with a series of natural disasters such as the Wenchuan earthquake in 2008 and the expenditure of preparing for the 2008 Beijing Olympic Games, the Chinese government proposed the "Four Trillion Plan" at the end of 2008[2]. There was an unprecedented increase in fiscal spending and debt issuance, and the

---

[2] Premier Wen Jiabao proposed the "Four Trillion Plan" at the executive meeting of the State Council in November 2008. Four trillion

public debt ratio rose to 34 percent, with its growth rate as high as nearly 8 percent.

The third debt surge began in 2015. A credit boom swept China early that year with the emergence of many unregulated online financing platforms. New loans reached a record high of 11.72 trillion yuan in 2015 according to the People's Bank of China. In the same year, China's stock market also experienced sharp fluctuations. The Shanghai Composite Index rose continuously from around 3,000 points at the beginning of the year to a high of 5,178 in June. However, the bubble burst and the Shanghai index fell back to 3,000 by the end of September. Rising non-performing loans at banks, high corporate indebtedness and financial uncertainty also drag the deleveraging process. In 2016, in order to maintain financial market stability and stimulate the economy, the government expanded the deficit and adopted a series of measures, including liquidity support and guarantees, which led to a further increase in the public debt. The fourth debt surge began with the outbreak of COVID-19 in January 2020. China's GDP, for the first time, experienced negative growth in 2020Q1 due to the shutdown and blockade. The Shanghai Composite Index fell by 8% in a single day on February 3, 2020, with some bank's credit business almost stopped. On March 27, the Central Committee clarified to issue special Treasury bonds, a total of 1 trillion yuan, starting from mid-June. The increase in the government deficit and the sudden decline in economic growth caused a temporary surge in the public debt in 2020.

**(Insert Fig. 1 here)**

## 4 Empirical results

*4.1 Benchmark model*

In this section, we first use the FTP-MS to identify two regimes of China's public debt. We include different control variables to evaluate the identification of different benchmark models and present the results in Table 2. The results of M1 to M5 show that China's public debt has the feature of "high growth rate and high volatility" and "low growth rate and low volatility" under the two regimes. For brevity, we refer to the "high growth and high volatility" regime as the "surge regime" and the "low growth and low volatility" regime as the "steady regime". In addition, the coefficients

---

in spending will be invested by the end of 2010 through ten major measures in order to stimulate domestic demand, increase employment and contribute to rapid economic recovery.

show that there is no significant multi-collinearity between the control variables, confirming the accuracy and reliability of the estimations. Based on AIC and likelihood values, we choose M5 as the final benchmark model and add financial variables on top of it in the next sub-section. More specifically, as shown in M5, China's public debt increases at an average rate of 7.239% per quarter in the surge regime with a variance of 1.210 ($e^{0.191}$), whereas it increases at an average rate of 2.676% per quarter in the steady regime with a variance of 0.634 ($e^{-0.455}$). This indicates that in the surge regime, the growth rate of public debt is about three times that of the steady regime, and the volatility is about twice that of the steady regime.

**(Insert Table 2 here)**

Figure 2 shows the smooth transition probability of being in the surge regime of the public debt in M5, which directly indicates the historical period of the public debt surge. According to Filardo (1994) and Hamilton (1989), public debt is regarded to be in the surge regime when the probability is greater than 0.5, otherwise, it is regarded to be in the steady regime. Therefore, the dates of these surge periods are 2007Q1-2007Q4, 2009Q1-2009Q4, 2015Q4-2017Q2, and 2020Q1-2020Q4. The average duration of the surge regime is 4.139 quarters, whereas that of the steady regime is 17.625 quarters. These periods correspond to the major economic and financial events in China mentioned above, confirming that the model can accurately identify and characterize the evolution characteristics of China's public debt.

There are some common points and differences between our results and the previous literature. In the study of Owusu et al. (2022), the authors use the levels of public debt ratio as the threshold variable to identify the two regimes and the regime threshold is estimated to be about 60%. In the threshold model proposed by Luigi and Huber (2018), the authors also use the growth rate series of public debt as the endogenous variable, which is the same in our paper. However, Luigi and Huber (2018) show that the regime threshold is estimated to be 0%, which means that the so-called "high debt" and "low debt" regime defined by the author is actually an expansion regime and a contraction regime of the US public debt. On the other hand, our results show that the average growth rate of China's public debt is positive in both regimes, reflecting the feature of China's long-term sustained growth of public debt and distinct fiscal pattern. In addition, Luigi and Huber (2018) also confirm that, in their sample, external shocks and major events will make the US public debt switch to the "high debt" regime, leading to a higher growth rate. This finding is consistent with what we find in China's public debt.

**(Insert Fig. 2 here)**

*4.2 The impact of financial variables on public debt growth rate*

Next, we include three financial variables respectively in the benchmark model M5 while maintaining the fixed transition probability. Table 3 shows the results of the impact of the growth rate of credit, house prices, and stock prices on the growth rate of public debt under different regimes. First, according to M6 and M7, in the surge regime, the coefficients of credit and house prices are significantly negative (-0.395 and -0.314, respectively). This means that in the surge regime, a 1% increase in credit growth rate and house price growth rate reduces the growth rate of public debt by 0.395% and 0.314%, respectively. In the steady regime, the impacts of credit and house prices are insignificant. However, stock prices exhibit an opposite result. The result of M8 shows that the coefficient of stock prices is significantly negative (-0.010**) in the steady regime, but insignificant in the surge regime. In other words, a 1% increase in the stock price growth rate reduces the growth rate of public debt by 0.010%, whereas in the surge regime, the impact is not significant. In addition, the results also show that these financial variables do not show significant multicollinearity, and the features of the two regimes remain unchanged.

**(Insert Table 3 here)**

The results above confirm that the increase in the growth rate of financial variables can reduce the growth rate of public debt, and financial variables also serve as leading indicators in this regard. This is consistent with the literature emphasizing that booms in financial markets will increase fiscal revenue and reduce public debt through multiple real channels (e.g., Chen et al., 2020a; Kiyotaki and Moore, 1997; Tagkalakis, 2011, 2012). The novelty of this study is that we examine the role of financial variables in different public debt regimes based on the nonlinear framework, contributing to formulating appropriate financial policies more accurately under different debt conditions.

First, the bust and collapse of credit affect economic stability and restrict business activity, causing a downturn in the private sector and ultimately hindering long-term economic growth (Levine, 2005). In response, governments usually take great expenditures to restore the market, re-establish the credit flow to stimulate the economy, and aid industries that face borrowing constraints. Therefore, a higher credit growth implies a stronger increase in fund availability and active investment by the private sector, restoring market confidence, and repairing corporate balance sheets and the financial system (Aghion et al., 2014). It also reflects less demand for expansionary fiscal policies and increasing tax revenues, which will significantly reduce the pressure on government spending and the growth rate of public debt. In addition, the increase in credit during the crisis may also amplify the government expenditure multiplier simultaneously, making economic growth faster than the increase in the public debt ratio, and ultimately resulting in the slowdown of public debt growth (Bernardini and Peersman, 2018; Borsi, 2018; Kim, 2023).

Second, the result of the house price variable in the surge regime well-consists with China's reality. On the one hand, more than half of the household wealth is occupied by real estate assets in China[3] and the high house-income ratio also exacerbates the impact of house price changes on Chinese household net assets (Chen et al., 2020a). On the other hand, the industrialization and urbanization in China in the past few decades have been largely driven by the real estate industry. The unique land finance system and the "land-infrastructure-leverage" strategy (Tsui, 2011) make the real estate industry closely related to public debt growth, becoming an important driver of fiscal balance and regional economic development by affecting land transfer fees, property taxes and turnover taxes (Gibb and Hoesli, 2003; Zhang et al., 2016). Therefore, the decline in house price growth significantly reduces fiscal revenues, limits regional economic growth, and lowers investment and consumption through the "wealth effect" (Lu et al., 2020; Singh, 2022). In the surge regime, a higher house price growth helps to mitigate these losses, generate more tax revenue for the government, stabilize market expectations, reduce government spending during the crisis, and moderate the growth of public debt.

It is worth noting that both credit and house price growth have significant impacts in the surge regime, whereas the impacts are insignificant under the steady regime. It shows the close linkage between credit and real estate market development in China. For example, in addition to mortgage loans, China's commercial banks provide a large proportion of loans to real estate developers, resulting in a large exposure to the real estate market (Zhang et al., 2014). During crises, profits of the real estate market and mortgage asset values decline, non-performing loans increase, and the government may inject a large amount of money in response (Jiang et al., 2013; Su et al., 2021). In this regard, facing external shocks, the formulation of coordinated policies between credit and real estate markets can have better effects on reducing the deficit and public debt growth, e.g., priorly allocating credit flows to constrained housing demand. Besides, credit and real estate markets do not seem to be the key factors in public debt growth in the steady regime, which indirectly reflects that during that period, the Chinese government may not have effectively incorporated capital changes in the credit and real estate markets into the fiscal adjustment, or it may just continue to expenditure rather than building fiscal buffers.

---

[3] According to the data of "Ifeng.com Financial Yunfeng Conference in 2020", about 20% of Chinese household assets are financial assets, of which about 80% come from real assets, i.e., only less than 10% of financial assets are in the stock market (https://finance.ifeng.com). The China Family Financial Investigation and the Research Center of the Southwest University of Finance of China also show that most Chinese family assets are occupied by real estate, and only about 10% to 20% are occupied by financial assets (https://chfs.swufe.edu.cn).

Finally, in the steady regime, the increase in stock price growth activates market transactions and stimulates consumption through the "wealth effect" (Sonje et al., 2014), increases stamp duty tax[4] for the government, and brings profitable returns to the government investment fund[5]. These revenues reduce the burden on government finances, which in turn helps to reduce the growth rate of public debt. However, the results of the surge regime show that the impact of stock price growth on the growth rate of public debt is not critical during external shocks or crises. Governments usually aim to stabilize economic growth, restore credit, and secure jobs in the short term. Changes in stock prices seem to marginally impact government finance compared to the fiscal expenditure and cost during crises.

*4.3 The impact of financial variables on regime transition probability*

Financial variables also affect the persistence of the growth pattern of public debt, i.e., by affecting the transition probability between regimes and thereby prolonging or shortening the duration of different debt regimes. Based on benchmark model M5, we include financial variables in the transition probability equation of the model, thereby forming the TVTP-MS. Table 4 shows the impacts of credit growth, house price growth and stock price growth on the transition probability of the public debt regimes. The results indicate that $\alpha_1^1$ (-0.759) in M9 is significantly negative whereas $\alpha_1^2$ is not, which shows that an increase in credit growth rate reduces the probability of public debt staying in the surge regime (which can also be interpreted as increasing the probability of switching to the steady regime). When the public debt is already in the steady regime, the credit growth rate has no effect on the transition probability. Besides, $\alpha_1^2$ in both M10 and M11 are significantly positive (0.432 and 0.091) but $\alpha_1^2$ are not. It indicates that the increase in the growth rate of house prices and stock prices increases the probability of public debt staying in the steady regime (which can also be interpreted as reducing the probability of switching to the surge regime). When public debt is already in the surge regime, the growth rate of house prices and stock prices have no significant impact.

---

[4] The scale of transaction tax in China's stock market is in the leading place worldwide and is charged on both sides of the transaction. Although relevant taxes and fees have been lowered several times in recent years as the market develops and opens, this tax would still bring revenues to the fiscal account, see Baltagi et al. (2006) for details. See the latest "Stamp Tax Law of the People's Republic of China" from the State Taxation Administration of the People's Republic of China at https://www.chinatax.gov.cn/eng/c102441/c5212066/content.html.

[5] The Chinese government investment fund also plays an important role in the industry R&D and debt issuance, and see Shao et al. (2023) for details.

**(Insert Table 4 here)**

The results above, for the first time, verify that financial variables can affect the growth pattern of public debt by changing the transition probability of regimes. This new finding contributes to the existing literature and sheds light on the role financial markets play in public debt sustainability. The results also share some similarities with Poghosyan (2018) who finds busts of credit, house prices, and stock prices prolong the duration of public debt expansions. It demonstrates that changes in financial variables indeed have a strengthening and weakening effect on the current public debt phase. However, in our study, we present the results from the perspective of regime transition probability. In other words, when the probability of staying in (or switching to) a certain regime increases, it means that the duration of this regime extends (shortens). In addition, Claessens et al. (2012) find that the economic recession accompanied by the bust of asset prices would be prolonged based on the cyclical perspective. It shows that financial shocks and real economic shocks can be amplified by financial accelerators and related mechanisms, reflecting on prolonging or shortening the duration of a certain phase of the cycle. In terms of analysis of the mechanism, the channels of financial variables that impact the growth of public debt in this paper are also highly consistent with those proposed by Claessens et al. (2012), which confirms that the real economy is the intermediary of financial variables affecting the growth of public debt, and emphasizes the importance of the financial market in economic development and debt sustainability.

*4.4 Robustness check*

To check the robustness of the impact of financial variables on the growth of public debt, we include financial variables in the regression equation and the transition probability equations simultaneously. Table 5 shows that the results remain robust, and the feature of China's public debt regimes is unchanged. In summary, we confirm that financial variables, serving as leading indicators, have an impact on the growth rate of China's public debt and the regime transition probability. For brevity, the smooth transition probability plots of M6 to M14 are shown in Appendix A.

**(Insert Table 5 here)**

# 5 Conclusions

Based on the quarterly data from 1998Q1 to 2023Q2, this paper employs the TVTP-MS to identify the regimes of China's public debt and verifies that the financial market has a significant impact on the growth of public debt. According to the empirical results, the two regimes of China's public debt show the features of high growth rate and high volatility, i.e., the surge regime, and low growth and low volatility, i.e., the steady regime. The dates of the surge regime of public debt correspond to several major external shocks and crises in history. The impacts of different financial variables (credit, house prices, stock prices) on the growth of public debt are twofold. On the one hand, a higher growth rate in credit and house prices reduces the growth rate of public debt in the surge regime, whereas the impacts of credit and house prices are insignificant in the steady regime. In addition, an increase in the stock price growth rate reduces the growth rate of public debt in the steady regime but has no such impact in the surge regime. On the other hand, financial markets also affect the probability of public debt being in a certain regime. Specifically, a higher credit growth rate reduces the probability of public debt staying in the surge regime. However, when public debt is already in the steady regime, credit has no such effect. In addition, an increase in the growth rate of house prices and stock prices increases the probability of public debt staying in the steady regime. Likewise, when public debt is already in the surge regime, the impacts of house prices and stock prices are insignificant. Therefore, this paper confirms the heterogeneous impacts of different financial variables on public debt growth. More importantly, we demonstrate for the first time that financial markets can affect the transition probability of regimes where public debt is located, suggesting the need to adjust financial variables in accordance with public debt regimes while making sustainable fiscal policies.

Our findings bring some implications. Policymakers should recognize the non-negligible role of financial markets in fiscal adjustments and public debt management, and make financial policies in accordance with the current debt conditions to better control public debt growth. First, facing external shocks and crises, the government should primarily and immediately restore the credit flows and the vitality of credit markets to mitigate the adverse impact on public debt. Controlling credit growth in the steady regime is also needed as low credit levels help to repair corporate balance sheets and make conditions for faster recovery and lower public debt growth when facing shocks in the future. Second, the stable growth of house prices should be guaranteed at all times as a healthy real estate market is crucial to the sustainability of China's public debt (Dong et al., 2017). Given the close linkage between China's credit market and the real estate industry, the government can formulate coordinated policies for credit and real estate, e.g., enhancing the supervision of the quality of real estate loans, optimizing the loan structure of commercial banks in time when the economy goes well, prioritizing the allocation of credit flows to constrained housing demand loans and providing guarantees priorly for real estate-related loans during crisis. Finally, the government should also promote the development of the stock market since the rise of stock prices brings considerable revenue for the government to build sufficient fiscal buffers and release debt space, which helps to ensure its sufficient fiscal capacity to cope with future economic uncertainties. Future studies can be deepened in several directions. First, with more sub-sector credit data of China in the future, it would be possible to study the impact of different types of credit on public debt growth. Second, since the sample of this paper is China's total public debt, it would be interesting to further explore the growth pattern and factors of central government debt and local government debt under different regimes.

**Declare of interest**

The authors declare no conflict of interest.

**Funding**

This work was supported by the Key Program of the National Natural Science Foundation of China [grant number 72033001]; the National Natural Science Foundation of China [grant number 72203019]

# Appendix A

Figure A.1 shows the smooth transition probability of public debt being in the surge regime identified by M6 to M14. According to Filardo (1994) and Hamilton (1989), when the smooth transition probability is larger than 0.5, public debt is considered to be in the surge regime, otherwise, it is considered to be in the steady regime. The periods corresponding to the shaded areas are several major external events in China.

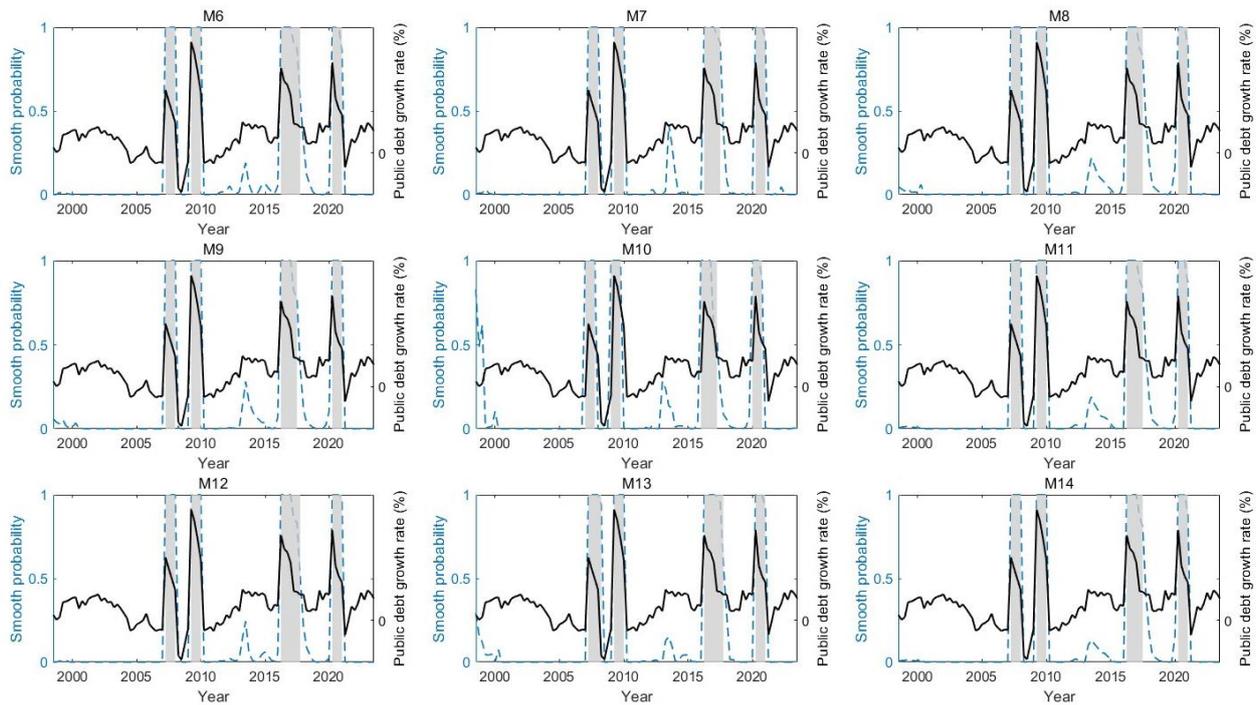

**Fig. A1. The smooth transition probability of the public debt regime identified by M6 to M14: The solid black line is the growth rate of public debt (right axis), the dashed blue line is the smooth transition probability of public debt being in the surge regime (left axis), and the shaded area is the period of public debt being in the surge regime.**

# Figures

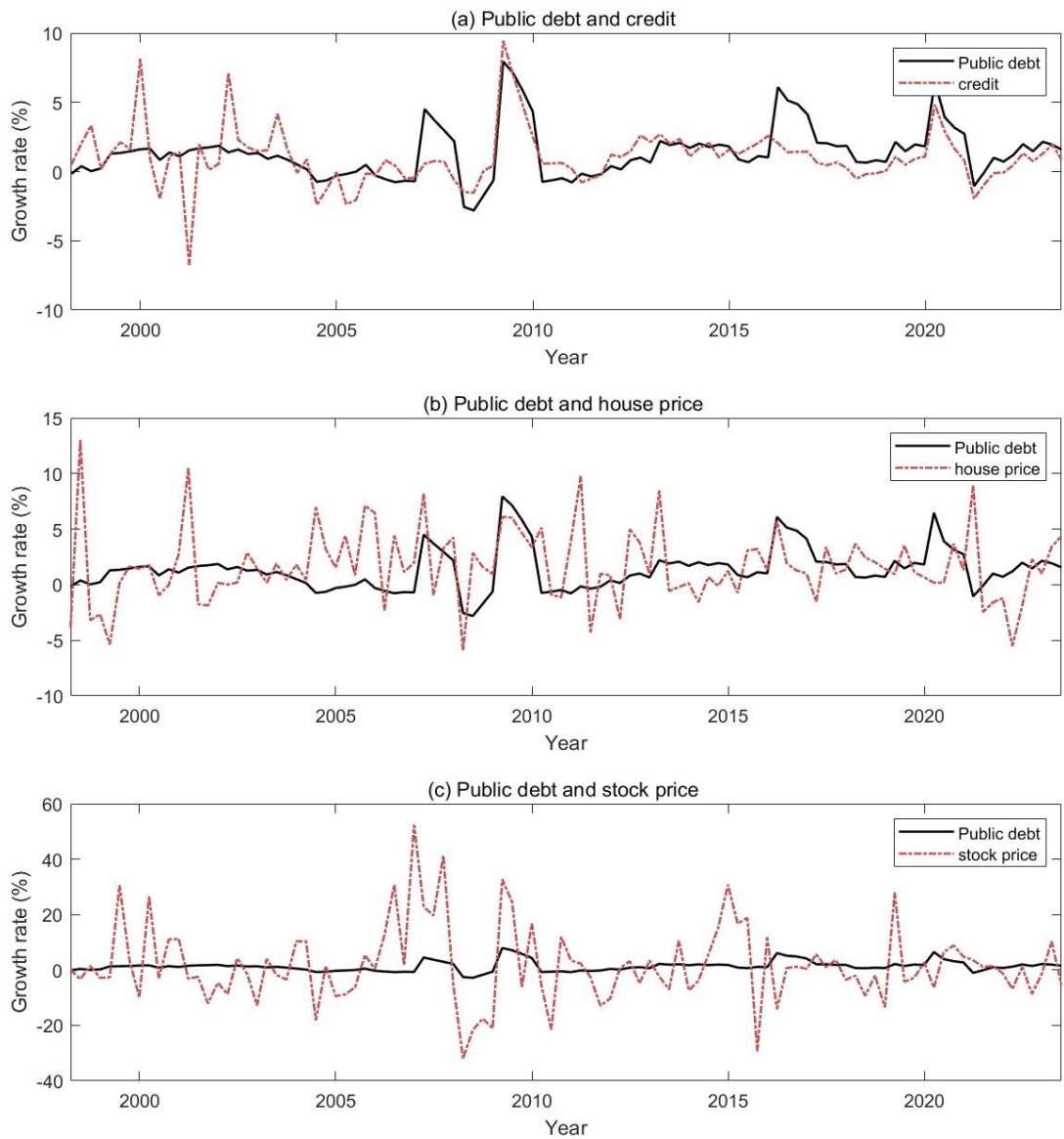

**Fig. 1. Time series plots of the growth rate of public debt and financial variables:** The black solid line is the growth rate of public debt, and the red dotted line is the growth rate of credit, house price, and stock price respectively.

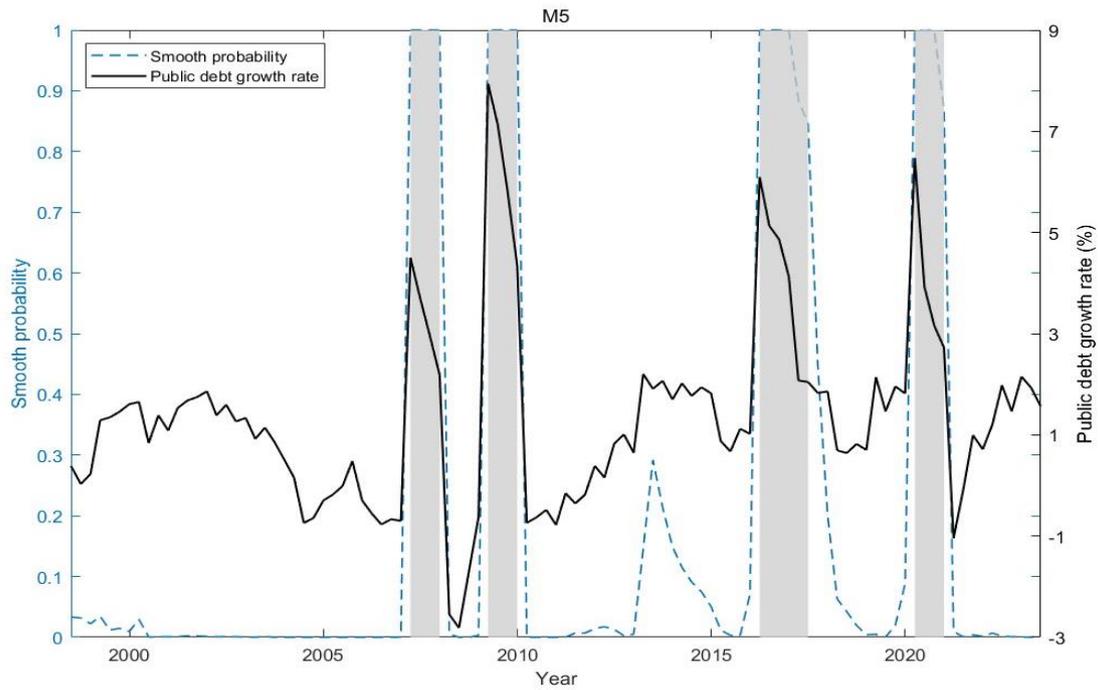

**Fig. 2. Smooth transition probability of public debt being in the surge regime in M5:** The solid black line is the growth rate of public debt (right axis), the dashed blue line is the smooth transition probability of public debt being in the surge regime (left axis), and the shaded area is the period of public debt in the surge regime.

# Tables

**Table 1. Basic statistics of the variables**

| Variable | Mean value | Stander deviation | Maximum | Minimum | Unit root test (*t*-statistics) | Number of observations |
|---|---|---|---|---|---|---|
| *Public debt and financial variables* | | | | | | |
| $pd$ | 1.351 | 1.875 | 7.942 | -2.814 | -3.942** | 102 |
| $credit$ | 1.054 | 2.080 | 9.442 | -6.755 | -5.454*** | 102 |
| $house$ | 1.666 | 3.350 | 13.020 | -5.935 | -5.974*** | 102 |
| $stock$ | 1.799 | 13.928 | 52.475 | -31.959 | -4.825*** | 102 |
| *Control variables* | | | | | | |
| $y$ | 2.467 | 2.687 | 11.936 | -12.202 | -8.755*** | 102 |
| $ca$ | 3.144 | 2.665 | 11.080 | -1.060 | -1.893 | 102 |
| $r$ | 3.321 | 1.238 | 8.666 | 1.080 | -4.908*** | 102 |
| $oil$ | 3.584 | 20.313 | 91.748 | -66.459 | -11.906*** | 102 |

Note: All the results in this table are presented in percentages. $pd$, $credit$, $house$, $stock$, $y$ and $oil$ refer to the growth rate (in percentages) of the public debt ratio, credit ratio, house price, stock price, IAV and oil price respectively. $ca$ and $r$ refer to the current account balance-to-GDP ratio and the interest rate. The column "Unit root test" presents the *t*-statistics value of the ADF unit root test (Null hypothesis: the series has a unit root). ** and *** represent significance at 5% and 1% level.

**Table 2. Results of the benchmark model**

| | Dependent variable: the growth rate of public debt | | | | |
|---|---|---|---|---|---|
| | M1 | M2 | M3 | M4 | M5 |
| ***Surge regime*** | | | | | |
| $\mu^1$ | 3.583*** | 4.845*** | 3.865*** | 1.590*** | 7.239*** |
| | (0.688) | (0.648) | (1.429) | (0.468) | (0.746) |
| $\log(\sigma_t^1)$ | 1.009*** | 0.490** | 0.972*** | 0.951*** | 0.191 |
| | (0.148 | (0.187) | (0.091) | (0.084) | (0.198) |
| $y_{t-1}^1$ | -0.257* | | | | -0.159** |
| | (0.148) | | | | (0.096) |
| $ca_{t-1}^1$ | | -0.036 | | | 0.049 |
| | | (0.119) | | | (0.094 |
| $r_{t-1}^1$ | | | -0.868** | | -1.044*** |
| | | | (0.455) | | (0.213) |
| $oil_{t-2}^1$ | | | | -0.036** | -0.016* |
| | | | | (0.020) | (0.007) |
| ***Steady regime*** | | | | | |
| $\mu^2$ | 1.536*** | 1.737*** | 1.855*** | 1.331** | 2.676*** |
| | (0.177) | (0.139) | (0.165) | (0.124) | (0.272) |
| $\log(\sigma_t^2)$ | -0.296** | -0.241*** | -0.619*** | -0.581*** | -0.455*** |
| | (0.064) | (0.076) | (0.089) | (0.108) | (0.075) |
| $y_{t-1}^2$ | -0.271*** | | | | -0.276*** |
| | (0.062) | | | | (0.049) |
| $ca_{t-1}^2$ | | -0.349*** | | | -0.273*** |
| | | (0.037) | | | (0.033) |
| $r_{t-1}^2$ | | | -0.141** | | -0.146** |
| | | | (0.060) | | (0.061) |
| $oil_{t-2}^2$ | | | | 0.002 | 0.005 |
| | | | | (0.004) | (0.005) |
| AIC | 3.347 | 3.205 | 3.467 | 3.503 | 2.906 |
| Loglikelihood | -161.024 | -153.866 | -167.098 | -168.922 | -132.753 |
| Number of observations | 101 | 101 | 101 | 101 | 101 |

Note: The coefficients in this table are estimated from E.q. (1) with standard error in parentheses. The transition probability part of the model is based on E.q. (4) and E.q. (5), i.e., the FTP-MS. *, ** and *** represent significance at 10%, 5% and 1% level.

**Table 3. The impact of financial variables on the growth rate of public debt under different regimes.**

| | Dependent variable: the growth rate of public debt | | |
|---|---|---|---|
| | M6 | M7 | M8 |
| **Surge regime** | | | |
| $\mu^1$ | 11.104*** | 8.017*** | 6.880*** |
| | (2.554) | (1.277) | (0.862) |
| $\log(\sigma_t^1)$ | 0.095 | -0.032 | 0.166 |
| | (0.213) | (0.236) | (0.184) |
| $y_{t-1}^1$ | -0.221*** | -0.146*** | -0.140*** |
| | (0.104) | (0.081) | (0.899) |
| $ca_{t-1}^1$ | -0.003 | -0.026 | 0.078 |
| | (0.094) | (0.104) | (0.105) |
| $r_{t-1}^1$ | -1.975*** | -1.0392*** | -0.953*** |
| | (0.563) | (0.362) | (0.239) |
| $oil_{t-2}^1$ | -0.019** | -0.015** | -0.014* |
| | (0.008) | (0.007) | (0.007) |
| $credit_{t-1}^1$ | -0.395* | | |
| | (0.177) | | |
| $house_{t-4}^1$ | | -0.314*** | |
| | | (0.147) | |
| $stock_{t-3}^1$ | | | -0.016 |
| | | | (0.014) |
| **Steady regime** | | | |
| $\mu^2$ | 2.569*** | 2.714*** | 2.617** |
| | (0.300) | (0.292) | (0.253) |
| $\log(\sigma_t^2)$ | -0.466*** | -0.477*** | -0.417*** |
| | (0.074) | (0.083) | (0.078) |
| $y_{t-1}^2$ | -0.263*** | -0.258*** | -0.261*** |
| | (0.050) | (0.051) | (0.048) |
| $ca_{t-1}^2$ | -0.265*** | -0.271*** | -0.274*** |
| | (0.032) | (0.035) | (0.032) |
| $r_{t-1}^2$ | -0.141** | -0.157** | -0.132** |
| | (0.053) | (0.069) | (0.058) |
| $oil_{t-2}^2$ | 0.005 | 0.003 | 0.003 |
| | (0.005) | (0.005) | (0.005) |
| $credit_{t-1}^2$ | 0.057 | | |
| | (0.053) | | |
| $house_{t-4}^2$ | | -0.029 | |
| | | (0.024) | |
| $stock_{t-3}^2$ | | | -0.010** |
| | | | (0.005) |
| AIC | 2.893 | 2.847 | 2.905 |
| Loglikelihood | -130.130 | -127.798 | -130.730 |
| Number of observations | 101 | 101 | 101 |

Note: This table reports the results of the impact of financial variables on the growth rate of public debt under different regimes. The coefficients in this table are estimated from E.q. (8) with standard error in parentheses. The transition probability part of the model is based on E.q. (4) and E.q. (5), i.e., the FTP-MS. *, ** and *** represent significance at 10%, 5% and 1% level.

**Table 4. The impacts of financial variables on the transition probability of the public debt regimes.**

| | Dependent variable: the growth rate of public debt | | |
|---|---|---|---|
| | M9 | M10 | M11 |
| | $fin_{t-l_0} = credit_{t-4}$ | $fin_{t-l_0} = house_{t-4}$ | $fin_{t-l_0} = stock_{t-2}$ |
| **Surge regime** | | | |
| $\mu^1$ | 7.178*** | 6.982*** | 7.279*** |
| | (0.766) | (0.910) | (0.741) |
| $log(\sigma_t^1)$ | 0.190 | 0.167 | 0.185 |
| | (0.191) | (0.181) | (0.196) |
| **Steady regime** | | | |
| $\mu^2$ | 2.670*** | 2.661*** | 2.669*** |
| | (0.272) | (0.323) | (0.274) |
| $log(\sigma_t^2)$ | -0.454*** | -0.442*** | -0.451*** |
| | (0.073) | (0.075) | (0.075) |
| **Coefficients of transition probability** | | | |
| $\alpha_0^1$ | 2.394*** | 2.170 | 1.243** |
| | (0.826) | (2.723) | (0.637) |
| $\alpha_1^1$ | -0.759** | -0.364 | -0.004 |
| | (0.324) | (0.681) | (0.030) |
| $\alpha_0^2$ | 2.778*** | 2.941 | 3.327*** |
| | (0.573) | (0.655) | (0.784) |
| $\alpha_1^2$ | 0.111 | 0.432** | 0.091* |
| | (0.164) | (0.223) | (0.051) |
| AIC | 2.889 | 2.905 | 2.903 |
| Loglikelihood | -129.912 | -130.722 | -130.604 |
| Number of observations | 101 | 101 | 101 |

Note: This table reports the impacts of credit growth, the house price growth and the stock price growth on the transition probability of the public debt regimes. The coefficients of transition probability in this table are estimated from E.q. (10) and E.q. (11) with standard error in parentheses, i.e., the TVTP-MS. The regression part of the

model is based on E.q. (1).    *, ** and *** represent significance at 10%, 5% and 1% level.

**Table 5. Robustness check for the impact of financial variables on the growth of public debt.**

| | Dependent variable: the growth rate of public debt | | |
|---|---|---|---|
| | M12 | M13 | M14 |
| | Robustness check for the credit growth | Robustness check for the house price growth | Robustness check for the stock price growth |
| **Surge regime** | | | |
| $\mu^1$ | 11.127*** | 8.739*** | 6.891*** |
| | (3.492) | (0.722) | (0.927) |
| $\log(\sigma_t^1)$ | 0.097 | 0.019 | (0.160) |
| | (0.240) | (0.195) | |
| $credit_{t-1}^1$ | -0.398* | | |
| | (0.239) | | |
| $house_{t-4}^1$ | | -0.307*** | |
| | | (0.090) | |
| $stock_{t-3}^1$ | | | -0.016 |
| | | | (0.181) |
| **Steady regime** | | | |
| $\mu^2$ | 2.565*** | 2.599*** | 2.611*** |
| | (0.308) | (0.291) | (0.255) |
| $\log(\sigma_t^2)$ | -0.468*** | -0.473*** | -0.467*** |
| | (0.088) | (0.077) | (0.079) |
| $credit_{t-1}^2$ | 0.055 | | |
| | (0.043) | | |
| $house_{t-4}^2$ | | -0.022 | |
| | | (0.024) | |
| $stock_{t-3}^2$ | | | -0.010** |
| | | | (0.005) |
| **Coefficients for transition probability** | | | |
| $\alpha_1^1$ | -0.738*** | 0.333 | -0.004 |

|  | (0.294) | (0.083) | (0.031) |
|---|---|---|---|
| $\alpha_1^2$ | 0.116 | 0.390 | 0.093* |
|  | (0.281) | (0.244) | (0.052) |
| AIC | 2.877 | 2.857 | 2.902 |
| Loglikelihood | -127.330 | -126.325 | -128.573 |

Note: The coefficients of the regression part of the model are estimated from E.q. (8) in this table and the coefficients of the transition probability part of the model are estimated from E.q. (10) and E.q. (11), i.e., TVTP-MS. Standard errors are in parentheses. *, ** and *** represent significance at 10%, 5% and 1% level.